\documentclass[aps,preprint,epsfig,rotate]{revtex4}
\usepackage{graphicx}
\usepackage{bm}
\usepackage{epsfig}

\input epsf
\begin{document}
\title{Internal clusterization in the Coulomb few-body systems and stability of the five- and six-body systems with unit charges}

\author{Alexei M. Frolov}
\email[E--mail address: ]{alex1975frol@gmail.com}

\affiliation{Department of Applied Mathematics \\
       University of Western Ontario, London, Ontario N6H 5B7, Canada} 

\date{\today}

\begin{abstract}
Stability of the bound ground states in the six five-body muonic $a b \mu e_2$ ions (or $(a b \mu e_2)^{-}$) and ten six-body $a b c \mu e_2$ quasi-molecules
is investigated. The notations $a, b, c$ stand for the bare nuclei of three hydrogen isotopes - protium ($p$), deuterium ($d$) and tritium ($t$). All these 
systems are the Coulomb five- and six-body systems with unit charges. It is found that the ground bound $S(L = 0)-$states are stable in all six five-body 
$a b \mu e_2$ ions and in ten six-body $a b c \mu e_2$ quasi-molecules. As follows from accurate variational computations of the five-body $a b \mu e_2$ ions 
their internal structure is similar to the internal structure of the negatively charged hydrogen ion H$^{-}$. Analogously, we show that the internal structures 
of the six-body neutral $a b c \mu e_2$ quasi-molecules are similar to the structure of the two-center H$_{2}$ molecule which contains two heavy hydrogen nuclei 
and two bound electrons. 


\end{abstract}

\maketitle
\newpage

\section{Introduction}

In this study we investigate the bound state spectra of a number of Coulomb five- and six-body systems with unit charges. As is well known many Coulomb few-body
systems with unit electrical charges are of great interest in a large number of physical and astrophysical problems (see, e.g., \cite{Advc}, \cite{Drach1} and 
references therein). Therefore, it is important to develop the general theory of bound states in such systems. Note that the complete theory of bound state 
spectra in the Coulomb three-body systems with unit charges was created 25 - 30 years ago (see, e.g., \cite{Posh}, \cite{Fro92}). For analogous four-body 
systems we still do not have any complete theory of bound states, but for some classes in such systems, e.g., for the symmetric four-body systems $a^{+} a^{+} 
b^{-} b^{-}$ and $a^{+} b^{+} c^{-} c^{-}$ we can now make a number of accurate predictions about their stability, bound state spectra and some properties. For 
the five- and six-body Coulomb systems with unit charges the situation is substantially worse, since for such systems only a few accurate results of bound state 
numerical computations can be found in modern literature. Very likely, the boundness of any state in the five- and six-body systems with unit charges should be 
considered as some irregularity. Indeed, a large number of such systems are not bound et al (not even weakly-bound). For instance, in our earlier study 
\cite{Fro2011} we have shown that the five-body Ps$_2 e^{-}$ and Ps$_2 e^{+}$ ions are not bound. However, these ions become weakly-bound, if the outer-most 
electron in the Ps$_2 e^{-}$ ion can be considered as a different particle (or particle of `different symmetry') from the two internal electrons (electrons 
included in the central Ps$_2$ cluster). This can be the case, if the Ps$_2 e^{-}$ and/or Ps$_2 e^{+}$ ions are placed in some relatively strong electric or/and 
magnetic field which substantially affects the motion of the outer-most electron(s) only. A very similar situation can be found in the six-body Ps$_3$ (or 
$(e^{-} e^{+})_3$) system which is also known as tri-positronium \cite{Fro2011}. In general, to discuss such systems we need to define the `distance of actual 
indistigishability' which can vary in different experimental conditions. In this study, however, we cannot discuss this interesting problem.  

Nevertheless, there are some five- and six-body Coulomb systems with unit electrical charges which are certainly bound and even well-bound. Formation and stability 
of such systems is based on the clusterization effect known from numerous experiments and observations. For some five- and six-body Coulomb systems with unit 
charges this effect is crucial, since it provides the actual stability of a number of low-lying bound states. To explain clusterization in some five- and six-body 
systems with unit electrical charges let us consider the five-body ion $p d \mu e_2$ which contains two heavy particles (or atomic nuclei) $p$ (proton) and $d$ 
(deuteron), one negatively charged muon $\mu^{-}$ and two electrons $e^{-}$. First, we can predict the internal structure of this five-body ion. It is very likely, 
that in such an ion the two nuclei $p^{+}$ and $d^{+}$ can be bound together by one heavy negatively charged muon $\mu^{-}$. The spatial radius of the arising 
three-body muonic ion $(p d \mu)^{+}$ is very short $a_{\mu} \approx a_0 \Bigl(\frac{m_e}{m_{\mu}}\Bigr) \approx \frac{a_0}{207.7}$, where $a_0$ is the Bohr radius 
(radius of the $1^{2}s-$electron orbit in the hydrogen atom), $m_e$ is the electron mass, while $m_{\mu}$ is the muon mass. This value of $a_{\mu}$ is $\approx$ 
1.5 times smaller than the Compton wavelength of electron $\Lambda_e = \alpha a_0 \approx \frac{a_0}{137}$, where $\alpha$ is the fine structure constant. In other 
words, the arising three-body cluster $(p d \mu)^{+}$ is a very compact, positively charged and very heavy system. This means that the three-body $(p d \mu)^{+}$ 
ion (i.e. three-body cluster with unit charges) can play the role of positively charged quasi-nucleus, which is an almost point particle, and can bind one or two 
electrons $e^{-}$. 

The arising four- and five-body quasi-atomic systems can be considered as four-body quasi-atoms (e.g., $p d \mu e$ and $d d \mu e$), or five-body quasi-ions (e.g., 
$p d \mu e_2$ and $d t \mu e_2$). Moreover, such four-body `atoms' and five-body `ions' can interact with another hydrogen H and/or deuterium D atom, or with the 
hydrogen/deuterium molecule H$_2$ (or D$_2$). Such an interaction may lead to the formation of the six-body neutral quasi-molecules, e.g., $(p d \mu) p e_2, (d d 
\mu) p e_2, (d t \mu) d e_2$, etc. Below we shall designate these systems by using the following notations: $p p d \mu e_2, p d d \mu e_2, d d t \mu e_2$, etc. 
Note that all these systems contain only particles with unit charges. Formation of analogous seven- and eight-body quasi-molecular systems with unit charges is 
also possible, but in this study we restrict ourselves to the analysis of the five-body quasi-atomic ions, such as $p d \mu e_2, d d \mu e_2, d t \mu e_2$, etc, 
and six-body neutral quasi-molecules, such as $p d t \mu e_2, p d d \mu e_2, t t t \mu e_2$, etc. 

Based on our current knowledge of internal structures of various few-body atoms and ions we can predict that all five-body, two-electron $(a b \mu e_2)^{-}$ ions, 
where $(a, b) = (p, d, t)$, are similar to the negatively charged hydrogen ion H$^{-}$, which has only one stable electronic $1^1S$-state (or the ground (bound) 
state). Here and everywhere below the letters $a, b$ and $c$ designate the nuclei of hydrogen isotopes. In contrast with the five-body ions, the internal structure 
of each of the six-body quasi-molecules ($p d t \mu e_2, d d t \mu e_2$, etc) is similar to the structure of the two-electron and two-center H$_{2}$ molecule which 
is the four-body neutral Coulomb system with unit charges. Accurate numerical coincidence of the bound state properties determined from direct computations of the 
$a b \mu e_2$ and $a b c \mu e_2$ systems with the known expectation values of the H$^{-}$ ion and H$_{2}$ molecule, respectively, is the crucial test for our method. 
On the other hand, the muonic properties determined for these five- and six-body systems must coincide well with the analogous bound state properties of the three-body 
$a b \mu$ and $b c \mu$ ions (see, e.g., see, e.g., \cite{Fro2011} and references therein). Below, based on the results of numerical computations we show that these 
suggestions about the internal structure of these Coulomb five- and six-body systems $a b \mu e_2$ and $a b c \mu e_2$ are correct.   

Our goal in this study is to perform variational computations of the ground states in the five-body ions $a b \mu e_2 = (a b \mu e_2)^{-}$, where $(a, b) = (p, d, 
t)$, and in six-body $a b c \mu e_2$ quasi-molecules, where $(a, b, c) = (p, d, t)$. Briefly, we consider the bound states in a number of five- and six-body systems 
each of which contains one negatively charged muon $\mu^{-}$, two electrons $e^{-}$ and two (or three) bare nuclei of hydrogen isotopes, i.e. the nuclei of protium 
$p$, deuterium $d$ and/or tritium $t$. In our computations in this study we determine the total energies and basic geometrical and physical properties of these ions 
and quasi-molecules. All calculations in this work have been performed with the use the following particle masses:
\begin{eqnarray}
  & &m_{\mu} = 206.768262 m_e \; \; \; , \; \; \;  M_p = 1836.152701 m_e \label{eq01} \\
  & &M_d = 3670.483014 m_e \; \; \; , \; \; \; M_t = 5496.92158  m_e \nonumber
\end{eqnarray}
where $m_e$ is the electron's mass, while $m_{\mu}, M_p, M_d$ and $M_t$ are the masses of the $\mu^{-}$ muon, proton, deuteron and triton, respectively. The same set 
of particle masses was used in our earlier calculations of the three-body muonic molecular ions \cite{Fro2011}. In this study we apply the atomic units in which 
$\hbar = 1, e = 1$ and $m_e = 1$. 

By using these mass values we determine accurate solutions of the non-relativistic Schr\"{o}dinger equation $H \Psi = E \Psi$ for the discrete spectrum of a few-body 
system. The total energy $E$ must always be negative, i.e. $E < 0$, and such a systems must be bound (see below). The non-relativistic Hamiltonians of the five- and 
six-body systems considered in this study are
\begin{eqnarray}
  H_5 = -\frac{\hbar^{2}}{2 m_e} \Bigl[\nabla^2_{1} + \nabla^2_{2} + \frac{m_e}{m_{\mu}} \nabla^2_{\mu} + \frac{m_e}{M_a} \nabla^2_{a} + \frac{m_e}{M_b} \nabla^2_{b} 
 \Bigr] + \sum^{5}_{i=2} \sum^{4}_{j=1 (j<i)} \frac{q_i q_j e^{2}}{r_{ij}} \; \; \; , \; \; \label{Hamilt5}
\end{eqnarray}
and 
\begin{eqnarray}
  H_6 = -\frac{\hbar^{2}}{2 m_e} \Bigl[\nabla^2_{1} + \nabla^2_{2} + \frac{m_e}{m_{\mu}} \nabla^2_{\mu}  + \frac{m_e}{M_a} \nabla^2_{a} + \frac{m_e}{M_b} \nabla^2_{b} + 
 \frac{m_e}{M_c} \nabla^2_{c} \Bigr] + \sum^{6}_{i=2} \sum^{5}_{j=1 (j<i)} \frac{q_i q_j e^{2}}{r_{ij}} \; \; \; , \; \; \label{Hamilt6}
\end{eqnarray}
respectively. Here and everywhere below in this study we assume that the indeces 1 and 2 designate the two bound electrons, the index 3 stands for the negatively charged 
muon $\mu^{-}$, while indeces 4, 5 and 6 denote the two (or three) heavy hydrogen nuclei. In atomic units the explicit forms of these two Hamiltonains, Eqs.(\ref{Hamilt5}) 
and (\ref{Hamilt6}), are simplified, since in atomic units in Eqs.(\ref{Hamilt5}) - (\ref{Hamilt6}) we have $e^2 = 1, \hbar^2 = 1$ and $m_e = 1$. 

In general, stability of some $(L,S)-$state in the five-body $a b \mu e_2$ ion means that its total energy $E$ (which is the eigenvalue of the $H_5$ Hamiltonian, 
Eq.(\ref{Hamilt5})) must be lower than the energy $E_{tr}$ of the following dissociation process: $a b \mu e_2 \rightarrow (a b \mu) e + e^{-}$. The energy $E_{tr}$ 
approximately equals (but always exceeds!) the $\overline{E}_{tr}$ value which is the treshold energy for the dissociation process: $a b \mu e_2 \rightarrow (a b \mu)^{+} 
+ e^{-} + e^{-}$. The $\overline{E}_{tr}$ energy equals (in atomic units) $\overline{E}_{tr} = E(a b \mu) + E({}^{\infty}$H$^{-})$, where $E(a b \mu)$ and 
$E({}^{\infty}$H$^{-})$ are the total energies of the ground states in the three-body $(a b \mu)^{+}$ ion and two-electron hydrogen ion ${}^{\infty}$H$^{-}$ 
($E({}^{\infty}$H$^{-} \approx$ -0.52775101654437719659095$\ldots$ $a.u.$). Analogously, for six-body $a b c \mu e_2$ systems stability of some (bound) state means that 
the total energy of this state (which is the eigenvalue of the $H_6$ Hamiltonian, Eq.(\ref{Hamilt6})) must be lower than the $E_{tr}$ energy of the following dissociation 
process: $a b c \mu e_2 \rightarrow (b c \mu) e + a e$, where $a$ is the nucleus of the lightest hydrogen isotope. The $E_{tr}$ energy approximately equals (but always 
exceeds!) the $\overline{E}_{tr}$ energy which corresponds to the following dissociation process: $a b c \mu e_2 \rightarrow (b c \mu)^{+} + e^{-} + a^{+} + e^{-}$. 
Numerical value of $\overline{E}_{tr}$ equals (in atomic units) $\overline{E}_{tr} = E(a b \mu) + 2 E({}^{\infty}$H), where $E(a b \mu)$ are the total energies of the 
ground states in the three-body $(a b \mu)^{+}$ ion and one-electron hydrogen atom ${}^{\infty}$H, i.e. $E({}^{\infty}$H) = -0.5 $a.u.$ 

This paper has the following structure. Variational computations of the five-body negatively charged ions $a b \mu e_2$, where $(a, b) = (p, d, t)$, are considered in the 
next Section. Here we compare some of the bound state properties of these ions with the known expectation values obtained earlier for the negatively charged hydrogen ion(s) 
H$^{-}$ \cite{Fro2015}. Numerical calculations of the total energies and a few other bound state properties of the six-body quasi-molecules $a b c \mu e_2$, where $(a, b, 
c) = (p, d, t)$, are performed in Section III. Comparison of these properties with the known properties of the two-electron (or four-body) ${}^{1}$H$_{2}$ molecule (or 
$p p e e$ four-body system) is also performed can also be found in that Section. Classifications of the bounds states in the five-body $a b \mu e_2$ ions and six-body 
$a b c \mu e_2$ quasi-molecules are explained in the fourth Section, while the concluding remarks can be found in the last Section. Applications of the five- and six-body 
$a b \mu e_2$ and $a b \mu e_2$ systems to actual physical problems are discussed in the Appendix.
    
\section{Negatively charged five-body muonic ions}

As mentioned above for each ground (bound) state in the three-body $a b \mu$ ion the energy spectrum of each of the five-body negatively charged ions $a b \mu e_2$ (where 
$(a, b) = (p, d, t)$) contains only one bound electron state, which is the ground $1^1S_{e}$-state, where the notation $e$ means the electron. Formally, the  `rotationally' 
and `vibrationally' excited states in the three-body $a b \mu$ ions are needed an additional consideration. Indeed, in actual applications it is important to 
know the total number of bound states in each of the $a b \mu e_2$ ions. To answer this question we need to investigate the internal structure of these ions. In general, 
the five-body negatively charged ion $a b \mu e_2$ includes a central, very heavy quasi-nucleus $a b \mu$ (or ($a b \mu)^{+}$) and two bound electrons. The spatial radius 
of the central nucleus is in $\Big(\frac{m_{\mu}}{m_e}\Bigl) \approx 206.768$ times smaller than the radius of $1s-$electron orbit in the neutral hydrogen atom. In other 
words, this quasi-nucleus is a very compact three-body system. Electromagnetic coupling (besides the direct Coulomb interactions) between the two outer electrons and 
particles which are included in the three-body muonic $a b \mu$ ion (quasi-nucleus) is very small. This means that such an interaction cannot change the actual electronic 
structure of the $a b \mu e_2$ ion. In other words, each bound state in the three-body muonic $(a b \mu)^{+}$ ion corresponds to the unique electronic bound state in the 
five-body negatively charged ion $a b \mu e_2$. |In general, note that each of the three-body muonic $(a b \mu)^{+}$ ions has a few different bound states (muonic bound 
states). 

For instance, each of the `protium'-muonic $pp\mu, pd\mu$ and $pt\mu$ ions has only two bound muonic states, while each of the `deuterium'-muonic $dd\mu$ and $dt\mu$ ions 
has five bound muonic states, etc (for more detail, see \cite{Fro2011}). The total energies of such `muonic' bound states vary between a few atomic units and a few dozens 
atomic units. The corresponding energies of electron bound states are $\approx$ 10 - 20 times smaller. Therefore, in experiments with the five- and six-body muonic systems 
$a b \mu e_2$ and $a b c \mu e_2$ we can observe a few different series of bound electronic states in each of these system. In general, each bound state in the 
three-particle muonic quasi-nucleus $(a b \mu)^{+}$ generates one separate series of bound electronic states in the $a b \mu e_2$ and $a b c \mu e_2$ systems. Below, such 
series of the bound electronic states, each of which is generated by one bound muonic state, are called the fundamental series of bound (electronic) states. The bound state 
spectra of the $a b \mu e_2$ ions are represented as combinations of a few fundamental series. In the lowest-order approximation there is no interference between the bound 
muonic states in the central $(a b \mu)^{+}$ quasi-nucleus and electronic bound states in the five-body $a b \mu e_2$ systems (see also a discussion below). It follows from 
here that in the five-body $a b \mu e_2$ ions each fundamental series of bound electronic states contains only one bound (ground) $1^1S_{e}-$state.   

To determine the accurate solutions of the non-relativistic Schr\"{o}dinger equation $H \Psi = E \Psi$ (where $E < 0$) for various five-body Coulomb systems we apply the 
variational expansion of the wave function $\Psi$ written in multi-dimensional gaussoids in the relative coordinates $r_{ij} = \mid {\bf r}_j - {\bf r}_k \mid = r_{kj}$ 
\cite{KT}, \cite{Fro2008}. Here, the symbol ${\bf r}_i$ designates the Cartesian coordinates of the $i-$th particle. Each of these gaussoids explicitly depends upon a 
complete set of the relative coordinates $r_{ij}$. Note that these relative coordinates $r_{ij}$ are rotationally and translationally invariant. This means that these 
coordinates do not changes at any rotation and/or translation of the whole few-body system in our three-dimensional space. Therefore, the center-of-mass of this few-body 
system is separated automatically. In turn, the Hamiltonian(s) of such systems (see, Eqs.(\ref{Hamilt5}) and (\ref{Hamilt6})) can be used in the original (i.e. Cartesian) 
coordinates. 

For the five-particle $a b \mu e_2$ ions one finds ten relative coordinates $r_{ij} = r_{12}, r_{13}, r_{14}, r_{15}, r_{23}, \ldots, r_{45}$. Formally, only nine of these 
relative coordinates are truly independent, but this fact does not complicate analytical operations and numerical computations in the basis of multi-dimensional gaussoids 
\cite{KT} (in contrast with the basis of linear exponents). For the ground $S(L = 0)-$state the variational wave function of the five-body $a b \mu e_2$ ion, where $a \ne b$, 
written in multi-dimensional gaussoids takes the form: 
\begin{eqnarray}
 \Psi(r_{12}, r_{13}, \ldots, r_{45}) = \sum^{N_A}_{i=1} C_i (1 + \hat{P}_{12}) \psi_i = \sum^{N_A}_{i=1} C_i (1 + \hat{P}_{12}) [\exp(-\sum_{(jk)} \alpha^{(i)}_{jk} 
 r^2_{jk})] \; \; \; \label{eq3}
\end{eqnarray}
where $N_A$ is the total number of basis functions $\psi_i(r_{12}, r_{13}, \ldots, r_{45})$ used, $C_i$ are the linear parameters of this variational expansion, while 
$\alpha^{(i)}_{jk}$ are the non-linear parameters which are varied in actual calculations. The internal sum of the exponent in Eq.(\ref{eq3}) is calculated over all 
different pairs of particles, i.e. $(jk) = (jk) = (12), (13), \ldots, (45)$. Note that the wave functions, Eq.(\ref{eq3}), corresponds to the spatial part of the total 
wave function, since it does not contain any spin functions. The trial wave function, Eq.(\ref{eq3}), has the electron-electron permutation symmetry which corresponds to 
the singlet two-electronic states, i.e. singlet permutation symmetry between particles 1 and 2 in our notation.

In the five-body $p p \mu e_2, d d \mu e_2$ and $t t \mu e_2$ ions we have an additional pair of indistinguishable particles. Antisymmetrization of the total wave functions 
for such ions is slightly complicated. For the spatial part of the total wave function of the $a a \mu e_2$ ion we can write in our notations:
\begin{eqnarray}
 & & \Psi(r_{12}, r_{13}, \ldots, r_{45}) = \sum^{N_A}_{i=1} C_i (1 + \hat{P}_{12}) (1 + \hat{P}_{45}) \psi_i = \sum^{N_A}_{i=1} C_i (1 + \hat{P}_{12}) (1 + \hat{P}_{45})
 \; \; \; \label{eq35} \\
 & & [\exp(-\sum_{(jk)} \alpha^{(i)}_{jk} r^2_{jk})] = \sum^{N_A}_{i=1} C_i (1 + \hat{P}_{12} + \hat{P}_{45} + \hat{P}_{12} \hat{P}_{45}) [\exp(-\sum_{(jk)} 
 \alpha^{(i)}_{jk} r^2_{jk})] \nonumber 
\end{eqnarray}
where $\Psi$ is the spatial part of the total wave function of this ion. The spin part of the total wave function of the $a b \mu e_2$ ion is, in fact, a spatial constant. 
This allows us to consider the spatial wave function $\Psi$ as the total wave function of the $a a \mu e_2$ ion. The basis wave function $\psi_{i}$ in Eq.(\ref{eq3}) is 
written in the form 
\begin{eqnarray}
 \psi_{i} = \exp(-\alpha^{(i)}_{12} r^2_{12} &-&\alpha^{(i)}_{13} r^2_{13} -\alpha^{(i)}_{14} r^2_{14} -\alpha^{(i)}_{15} r^2_{15} -\alpha^{(i)}_{23} r^2_{23}
 -\alpha^{(i)}_{24} r^2_{24} -\alpha^{(i)}_{25} r^2_{25} \; \; \; \label{eq4} \\ 
 &-&\alpha^{(i)}_{34} r^2_{34} -\alpha^{(i)}_{35} r^2_{35} -\alpha^{(i)}_{45} r^2_{45}) \nonumber
\end{eqnarray}
Correspondingly, the $\hat{P}_{12} \psi_{i}$ function takes the from:
\begin{eqnarray}
 \hat{P}_{12} \psi_{i} = \exp(-\alpha^{(i)}_{12} r^2_{12} &-&\alpha^{(i)}_{23} r^2_{13} -\alpha^{(i)}_{24} r^2_{14} -\alpha^{(i)}_{25} r^2_{15} -\alpha^{(i)}_{13} r^2_{23}
 -\alpha^{(i)}_{14} r^2_{24} -\alpha^{(i)}_{15} r^2_{25}  \; \; \; \label{eq45} \\
 &-&\alpha^{(i)}_{34} r^2_{34} -\alpha^{(i)}_{35} r^2_{35} -\alpha^{(i)}_{45} r^2_{45}) \nonumber
\end{eqnarray}
In other words, the $\hat{P}_{12} \psi_{i}$ function can be determined by replacing the corresponding non-linear parameters in the exponent of the $\psi_{i}$ basis function. The 
$\hat{P}_{45} \psi_{i}$ and $\hat{P}_{12} \hat{P}_{45} \psi_{i}$ functions are determined analogously:
\begin{eqnarray}
 \hat{P}_{45} \psi_{i} = \exp(-\alpha^{(i)}_{12} r^2_{12} &-&\alpha^{(i)}_{13} r^2_{13} -\alpha^{(i)}_{15} r^2_{14} -\alpha^{(i)}_{14} r^2_{15} -\alpha^{(i)}_{23} r^2_{23}
 -\alpha^{(i)}_{25} r^2_{24} -\alpha^{(i)}_{24} r^2_{25} \; \; \; \label{eq47} \\ 
 &-&\alpha^{(i)}_{35} r^2_{34} -\alpha^{(i)}_{34} r^2_{35} -\alpha^{(i)}_{45} r^2_{45}) \nonumber
\end{eqnarray}
and 
\begin{eqnarray}
 \hat{P}_{12} \hat{P}_{45} \psi_{i} = \exp(-\alpha^{(i)}_{12} r^2_{12} &-&\alpha^{(i)}_{23} r^2_{13} -\alpha^{(i)}_{25} r^2_{14} -\alpha^{(i)}_{24} r^2_{15} -\alpha^{(i)}_{13} r^2_{23}
 -\alpha^{(i)}_{15} r^2_{24} -\alpha^{(i)}_{14} r^2_{25} \; \; \; \label{eq49} \\ 
 &-&\alpha^{(i)}_{35} r^2_{34} -\alpha^{(i)}_{34} r^2_{35} -\alpha^{(i)}_{45} r^2_{45}) \nonumber
\end{eqnarray}
These formulas, Eqs.(\ref{eq4}) - (\ref{eq49}), allows one to obtain the basis functions with the correct permutation symmetry between all identical particles. The properly 
symmetrized basis functions. constructed above, can be used in variational calculations of the symmetric five-body $a a \mu e_2$ systems. 

By performing a careful optimization of all ten non-linear parameters in each basis function $\psi_{i}$ ($i$ = 1, $\ldots, N$), one finds an accurate approximation (if $N$ is relatively 
large) to the spatial part of the actual wave function of an arbitrary $a b \mu e_2$ ion, where $(a, b) = (p, d, t)$. Results of our variational calculations of the five-body negatively 
charged ions $a b \mu e_2$ (or $(a b \mu e_2)^{-}$ ions) can be found in Table I (in $a.u.$) which contains the total energies of all six five-body $a b \mu e_2$ ions. The overall accuracy 
of these calculations for five-body systems can approximately be evaluated as $1 - 2 \cdot 10^{-8}$ $a.u.$ (for the energies). As follows from direct comparison of the total energies from 
Tables I and II all six known five-particle $a b \mu e_2$ ions considered in this study are stable. A number of bound state properties of these ions (or expectation values) are presented 
in Table III (in atomic units). It is interesting to note that almost all electronic bound state properties of the five-body $a b \mu e_2$ ions coincide very well with the known expectation 
values of the negatively charged hydrogen ion(s) H$^{-}$ presented in Table IV. The expectation values of the H$^{-}$ ion have been determined with the use of our highly accurate variational 
expansion developed for three-body systems (it is described in detail in \cite{Fro2015}). 

As follows from Tables III and IV a large number of electron-nucleus, electron-electron and even nucleus-nucleus expectation values in the $a b \mu e_2$ ions coincide very well with the 
analogous expectation values determined for the two-electron (or three-body) H$^{-}$ ion to a good numerical accuracy. For the $\langle r_{ne} \rangle, \langle r^{2}_{ne} \rangle, \langle 
r_{ee} \rangle, \langle r^{2}_{ee} \rangle, \langle r_{nn} \rangle$ and $\langle r^{2}_{ee} \rangle$ expectation values the observed numerical coincidence is very good. This indicates 
clearly that the electronic structures of the Coulomb five-body ions $a b \mu e_2$ are similar to the electronic structure of the two-electron H$^{-}$ ion(s). On the other hand, the 
computed muonic properties of all six $a b \mu e_2$ ions are very close to the analogous muonic properties of the three-body $a b \mu$ (or $(a b \mu)^{+}$) ions (see, e.g., \cite{Fro2011} 
and references threin). All these facts unambigously show that our ideas about clusterization of the five-body $a b c \mu e_2$ ions are correct.  

\section{Six-body muonic quasi-molecules}

The neutral six-body systems with unit charges $p d t \mu e_2, d d t \mu e_2, d d d \mu e_2$ and $d t t \mu e_2$, etc, include three heavy, positively charged nuclei of hydrogen 
isotopes,one negatively charged muon $\mu^{-}$ and two electrons $e^{-}$. Note that each of these systems contains two heavy centers each of which is a positively charged particle 
(or quasi-particle). For instance, in the six-body $d d t \mu e_2$ system there are two quasi-nuclei with unit charges: (1) the three-body $dt\mu$ (or $(dt\mu)^{+}$) muonic ion, and (2) 
bare deuterium nucleus $d$ (or $d^{+}$). Three-body quasi-nuclei ($a b \mu$) are very compact and heavy positively charged `particle' which can bind one/two negatively charged electrons 
$e^{-}$. The arising six-body structure $d d t \mu e_2$ (or, in the general case, the $a b c \mu e_2$ systems) is a neutral quasi-molecular (or two-center) system with two bound 
electrons. $A$ $priori$ we can expect some similarity between the internal structures of these six-body muonic systems $a b c \mu e_2$ and two-electron (and two-center) hydrogen molecule 
H$_{2}$ (see below). To respect this fact, below we shall call these six-body systems the `quasi-molecules', i.e. molecular, two-electron structures with the two heavy Coulomb centers 
each of which have a positive electric charge (equals unity). 

As mentioned above the radius of the three-particle `muonic' quasi-nucleus $(dt\mu)^{+}$ (and analogous $(pd\mu)^{+}, (pt\mu)^{+}, (dd\mu)^{+}$, etc, quasi-nuclei) is $\approx 
\Bigl(\frac{m_{\mu}}{m_{e}}\Bigr) \approx 206.768$ times smaller than atomic radius of the hydrogen atom $a_0 = \frac{\hbar^2}{m_e e^2} \approx 5.292 \cdot 10^{-9}$ $cm$ (Bohr radius). 
Therefore, in the first approximation such three-body muonic $(a b \mu)^{+}$ ions can be considered as the two `point' (or sizeless) particles. To describe the six-body $a b c \mu e_2$ 
quasi-molecules, in this Section we shall use the same system of notation developed in previous Sections. In particular, the indeces 1 and 2 mean electrons $e^{-}$, the index 3 stands 
for the negatively charged muon $\mu^{-}$, while three other indeces (4, 5 and 6) designate three nuclei of hydrogen isotopes. Note that our notation $a b c \mu e_2$ used here contains 
some uncertainty, since the heaviest quasi-nucleus (or quasi-nucleus which includes the $\mu^{-}-$muon) can be either $a b \mu$, or $a c \mu$, or $b c \mu$. Below, we shall assume that 
such a quasi-nucleus always include the two heaviest nuclei of hydrogen isotopes. This means that the $p d t \mu e_2$ system contains the two quasi-nuclei: three-particle $(d t \mu)^{+}$ 
ion and one bare proton $p^{+}$. In other words, the $p d t \mu e_2$ notation does not mean that the six-body $p d t \mu e_2$ system include the three-body $(p t \mu)^{+}$ ion and a bare 
deuterium nucleus, or the three-body $(p d \mu)^{+}$ ion and a bare tritium nucleus. These two-electron systems with two other pairs of quasi-nuclei: $(p t \mu)^{+}$ + $d^{+}$ and 
$(p d \mu)^{+}$ + $t^{+}$ are also stable (or better to say `quasi-stable'), but their total energies are smaller than the total energy of the $p (d t) \mu e_2$ system. These quasi-stable 
systems are not considered in detail in this study, but they are mentioned in the next Section. In other words, our notation $a b c \mu e_2$ always means that the inequality $M_a \le M_b 
\le M_c$ is obeyed for three nuclear masses of the hydrogen isotopes $a, b, c$. 

The trial variational wave function of the the non-symmetric six-body $p d t \mu e_2$ quasi-molecule with the correct permutation symmetry is written in the form
\begin{eqnarray}
  \Psi = \sum^{N_A}_{i=1} C_i (1 + \hat{P}_{12}) [\exp(-\sum_{(jk)} \alpha^{(i)}_{jk} r^2_{jk})] \; \; \; \label{equ35}
\end{eqnarray}
while for the partially symmetric $p p d \mu e_2, p p t \mu e_2$ and $d d t \mu e_2$ quasi-molecules the wave function $\Psi$ takes the form
\begin{eqnarray}
  \Psi = \sum^{N_A}_{i=1} C_i (1 + \hat{P}_{12}) (1 + \hat{P}_{45}) [\exp(-\sum_{(jk)} \alpha^{(i)}_{jk} r^2_{jk})] \; \; \; \label{equ351}
\end{eqnarray}
For the partially symmetric $p d d \mu e_2, p t t \mu e_2$ and $d t t \mu e_2$  quasi-molecules the wave function $\Psi$ is written in a slightly different form (in our notation)
\begin{eqnarray}
  \Psi = \sum^{N_A}_{i=1} C_i (1 + \hat{P}_{12}) (1 + \hat{P}_{56}) [\exp(-\sum_{(jk)} \alpha^{(i)}_{jk} r^2_{jk})] \; \; \; \label{equ352}
\end{eqnarray}
In these equations the internal sum is taken over all 15 permutations of particles $(jk) = (kj) = (12), (13), (14), \ldots$, (45), (46) and (56). Again, we have to note that for 
an arbitrary six-body system one always finds fifteen relative coordinates $r_{ij}$, but in our three-dimensional space only twelve of them are truly independent. Nevertheless, 
as it was shown in \cite{KT} one can operate with the variational expansion, Eq.(\ref{equ351}), by assuming that all fifteen relative coordinates are independent of each other. 
This means that we can use the same formulas for the matrix elements and all variational parameters $\alpha^{(i)}_{jk}$ can be varied in calculations without any additional 
restriction. 

As follows from Eqs.(\ref{equ35}) - (\ref{equ352}) the explicit construction of the trial wave functions for the non-symmetric ($p d t \mu e_2$) and partially symmetric ($a a b 
\mu e_2$ and $a b b \mu e_2$) systems is relatively simple. However, for the $p p p \mu e_2, d d d \mu e_2$ and $t t t \mu e_2$ quasi-molecules which contain three identical 
hydrogen nuclei one finds a number of additional complications. For these systems the actual wave function cannot be represented as a single-term product of the spatial and spin 
functions. The correct wave functions of these six-body quasi-molecules must be represented as the properly symmetrized finite sums of products of the different spatial and spin 
functions. In general, there are the two independent spin functions for the system of three identical particles such as protons ($p$), deuterons ($d$) and tritons ($t$). To 
illustrate this fact let us consider the six-body $p p p \mu e_2$ quasi-molecule. The wave function of the $p p p \mu e_2$ system is represented in the following general form
\begin{eqnarray}
  \Psi &=& \sum^{N_A}_{i=1} C_i (1 + \hat{P}_{12}) {\cal A}_{456} \Bigl\{ [\exp(-\sum_{(jk)} \alpha^{(i)}_{jk} r^2_{jk})] \sum_n \phi^{(n)}_S(456) \Bigr\} \; \; \; \label{equ355} \\
    &=& \sum^{N_A}_{i=1} C_i (1 + \hat{P}_{12}) (1 - \hat{P}_{45} - \hat{P}_{46} - \hat{P}_{56} + \hat{P}_{456} + \hat{P}_{465}) 
   \Bigl\{ [\exp(-\sum_{(jk)} \alpha^{(i)}_{jk} r^2_{jk})] \sum_n \phi^{(n)}(456) \Bigr\} \nonumber 
\end{eqnarray}
where ${\cal A}_{456}$ is the complete antysymmetrizer for three identical particles with indexes (or numbers) 4, 5 and 6, i.e. ${\cal A}_{456} = 1 - \hat{P}_{45} - \hat{P}_{46} 
- \hat{P}_{56} + \hat{P}_{456} + \hat{P}_{465}$, where $\hat{P}_{jk}$ and $\hat{P}_{ijk}$ are the permutations of the two and three identical particles, respectively. The notation 
$\phi^{(n)}_S(456)$ stands for the spin function(s) of the three protons. Below we designate the spin functions of a single proton in the following  way: spin-up function is the 
$\alpha-$function, while spin-down function is the $\beta-$function (see, e.g., \cite{LLQ}). In these notations one finds the two following spin functions for the system of three 
protons: $\phi^{(1)}_S(456) = \alpha_4 \beta_5 \alpha_6 - \beta_4 \alpha_5 \alpha_6 = \alpha \beta \alpha - \beta \alpha \alpha$ and $\phi^{(2)}_S(456) = 2 \alpha_4 \alpha_5 \beta_6 
- \beta_4 \alpha_5 \alpha_6 - \alpha_4 \beta_5 \alpha_6 = 2 \alpha \alpha \beta - \beta \alpha \alpha - \alpha \beta \alpha$, where we number the particles by their location in 
formulas (briefly, it is called the `indexation by location'). It is clear that these two spin functions $\phi^{(1)}_S(456)$ and $\phi^{(2)}_S(456)$ are orthogonal to each other, 
i.e. they are independent of each other.  

To determine all matrix elements of the Hamiltonian and overlap matrixes we need to perform integration over spin variables of all identical particles, including two electrons 
(particles 1 and 2) and three protons (particles 4, 5, and 6). Details of this procedure and explicit formulas for the arising spatial projectors can be found, e.g., in \cite{Fro2011}. 
In calculations of the total energy and bound state properties of the $p p p \mu e_2, d d d \mu e_2$ and $t t t \mu e_2$ quasi-molecules we can apply only one spin functions, e.g., 
$\phi^{(1)}_S(456)$. The spatial projector for the particles 4, 5 and 6 in this case is ${\cal P}_{456} = \frac{1}{12} (2 + 2 \hat{P}_{45} - \hat{P}_{46} - \hat{P}_{56} - \hat{P}_{456} 
- \hat{P}_{546})$. In general, in numerical calculations of all regular (spin-indpendent) expectation values, including the total energy, one can apply one spin function only. This 
fact substantially simplifies all numerical computations. 

Results of our variational calculations of the total energies for all ten six-body $a b c \mu e_2$ quasi-molecules in their ground states can be found in Table I. The overall accuracy 
of these our calculations for six-body systems can be evaluated as $3 - 5 \cdot 10^{-8}$ $a.u.$ (for the energies). By comparing these energies with the data from Table II we conclude 
that the ground states in all ten quasi-molecules $p d t \mu e_2, p p p \mu e_2, d d d \mu e_2, t t t \mu e_2, p p d \mu e_2, p p t \mu e_2, p d d \mu e_2, p t t \mu e_2, d d t \mu e_2$ 
and $d t t \mu e_2$ are bound (or stable). A number of bound state properties of the six-body $p d t \mu e_2$ quasi-molecule in atomic units can be found in Table III. These expectation 
values must be compared with the expectation values determined for the ${}^{1}$H$_2$ molecule (see Table IV). Note that all expectation values for the ${}^{1}$H$_2$ molecule have been 
determind in this study with the use of the six-dimensional gaussoids which form the radial basis set of the four-body problem.  

As follows from the expectation values presented in Tables III and IV a large number of electron-nucleus, electron-electron and even nucleus-nucleus (or nucleus-quasi-nucleus) 
expectation values determined for the $p d t \mu e_2$ quasi-molecule coincide well with the analogous expectation values known for the two-electron ${}^{1}$H$_2$ molecule. For nine 
other $a b c \mu e_2$ quasi-molecules we observed analogous numerical coincidence of the bound state properties. Again, we want to emphazise an excelent agreement between the known 
properties of the three-body muonic ion $d t \mu$ \cite{Fro2011} and expectation values computed for the $p d t \mu e_2$ quasi-molecule. Results of such a comparison indicate that
our numerical computations of the bound states in the six-body $a b c \mu e_2$ quasi-molecules have been performed correctly and accurately. Furthermore, our results re-produce all 
essenetial details of internal structure of the six-body Coulomb systems $a b c \mu e_2$ with unit electrical charges. 

\section{Classification of the bound state spectra}

Let us discuss the possible classification scheme which can successfully be applied to describe the bound states in the five-body $a b \mu e_2$ ions and six-body $a b c \mu e_2$ 
quasi-molecules. Note that each of the five-body $a b \mu e_2$ ions and six-body $a b c \mu e_2$ quasi-molecules is an atomic/molecular two-electron system which contains a very 
compact three-body cluster with one negatively charged muon $\mu^{-}$, i.e. $p d \mu, p t \mu, d t \mu$, etc. As mentioned above the radius of such a compact, positively charged 
cluster $a b \mu$ (or $(a b \mu)^{+}$) is substantially smaller than the sizes of electronic orbits in the both five-body $a b \mu e_2$ ion and six-body $a b c \mu e_2$ 
quasi-molecule. As follows from here the three-body cluster $(a b \mu)^{+}$ will play the role of quasi-nucleus in these five-body $a b \mu e_2$ ions and six-body $a b c \mu e_2$ 
quasi-molecules. The total and binding energies of such a three-body quasi-nucleus $a b \mu$ are substantially larger (in dozens and hundreds times larger) than the corresponding 
atomic and molecular energies. Therefore, to classify the bound states in the whole five- and six-body muonic systems we need to mention (first of all) the conserving (or 
quasi-conserving) quantum numbers for the three-body $a b \mu$ cluster(s) and then the analogous quantum numbers for the whole six-body $a b c \mu e_2$ system. 

In general, the bound states in the three-body $a b \mu$ quasi-nucleus are classified by the `vibrational' $\nu$ and `rotational' $\ell$ quantum numbers $\nu$ and $\ell$. Formally, 
these two quantum numbers ($\nu$ and $\ell$) corresponds to the two-center, pure adiabatic three-body system, e.g., these quantum numbers are good for the three-body, one-electron) 
${}^{\infty}$H$_{2}^{+}$ ion which contains two infinite nuclear masses (centers). However, in actual muonic ions $p p \mu, p d \mu, \ldots, t t \mu$ these quantum numbers are not 
rigorously conserved. Nevertheless, the `rotational' $\ell$ and `vibrational' $\nu$ quantum number allows one unambigously to designate an arbitrary bound state in all three-body 
muonic ions $a b \mu$. 

To complete the classification of the bound state in the five-body $a b \mu e_2$ ions and six-body $a b c \mu e_2$ quasi-molecules one needs to add (to the $(\nu,\ell)_{\mu}$ 
notation) an additional set of `atomic' and `molecular' quantum numbers, which includes the `electronic' quantum numbers. It is clear that there is a fundamental 
difference between the one-center atomic sysytems (ion) and two-center quasi-molecules. For all one-center five-body $a b \mu e_2$ ions we can use the standard `atomic' quantum 
numbers $L$ and $S$ (see, e.g., \cite{LLQ} and \cite{Fock}). In this notation the ground state in the five-body $p t \mu e_2$ ion is designated as the $[(0, 0)_{\mu}; 1^{1}S_e]$ 
state. Here and below the index $\mu$ means muonic (bound) state, while the index $e$ designates electron bound state, or bound state of two electrons in such a quasi-atom. As 
mentioned above each of these five-body ions has the electronic structure similar to the electronic structure of the hydrogen negatively charged ion H$^{-}$, i.e. it has the only 
one stable bound $1^{1}S$-state (ground state).

Analogously, for the six-body $a b c \mu e_2$ quasi-molecule we need to add (to the $(\nu, \ell)_{\mu}-$notation) an additional notation which determines the bound state of the 
two-electron molecular systems with the two immovable centers (see, e.g., \cite{LLQ}, \cite{Eiry} and \cite{MQW}). It is clear that such a molecular system $a b c\mu e_2$ is similar 
to the regular H$_{2}$ (or HD) molecules. Therefore, the notation $[(0, 2)_{\mu}; ({}^{1}\Pi)_e]$ used for the $p d t \mu e_2$ system means the bound state in the six-body $p d t 
\mu e_2$ quasi-molecule, where the central muonic quasi-nucleus is in its bound rotationally excited $D(L = 2)$-state, while two electrons occupy the molecular ${}^{1}\Pi-$state. 
However, for the six-body $a b c \mu e_2$ quasi-molecule we need to add a few additional quantum numbers which are used to designate the rotational and vibrational states of the
whole two-center `molecule'. In general, we also need to indicate the both rotational $J(K)$ and vibrational quantum numbers $v$ of the whole two-center molecule (see, e.g., chapter 
11 in \cite{LLQ}) and/or the overall multiplicity (or spin-multiplicity) of the actual bound states of the both nuclei (bare nucleus $a$ and three-body quasi-nucleus $b c \mu$) and 
two electrons \cite{LLQ}. Here we do not want to discuss the complete notation in detail, since they are not needed in this study. 

The system of notations developed above can be used to describe the general structure of the bound state spectra of the five-body $(a b \mu e_2)^{-}$ ions and six-body $a b c \mu e_2$ 
quasi-molecules. First, consider the five-body $(a b \mu e_2)^{-}$ ions. The bound state spectrum of each of these (six) ions ($p p \mu e_2, p d \mu e_2, p t \mu e_2, d d \mu e_2, 
d t \mu e_2$ and $t t \mu e_2$) includes only one bound (electron) $1^{1}S_e-$state. In other words, all bound states in these five-body ions can be designated as the 
$[(\nu, \ell)_{\mu}; 1^{1}S_e]$-states, where $\nu$ and $\ell$ are the vibrational and rotational quantum numbers of the bound $(\nu, \ell)-$state of the central $a b \mu$ ion. As 
follows from here that each of the three light (or protium containing) $p b \mu e_2$ ions (where $b = p, d, t$) has only two stable (bound) states: (1) the ground 
$[(0,0)_{\mu}; (1^{1}S)_e]-$state, and (2) the excited $[(0,1)_{\mu}; (1^{1}S)_e]-$state. Analogously, the bound state spectra of each of the five-body $d d \mu e_2$ and $d t \mu e_2$ 
ions includes five bound states: the ground $[(0,0)_{\mu}; (1^{1}S)_e]-$state and four excited states: $[(0,1)_{\mu}; (1^{1}S)_e], [(0,2)_{\mu}; (1^{1}S)_e], [(1,0)_{\mu}; (1^{1}S)_e]$ 
and $[(1,1)_{\mu}; (1^{1}S)_e]$. The $[(1,1)_{\mu}; (1^{1}S)_e]$ state is a very weakly-bound state in each of these two ions, since, e.g., the binding energy of the (1,1)-state in the 
three-body $d t \mu$ ion is only $\approx$ 0.02435 \% of its total energy. The bound state spectrum of the five-body $t t \mu e_2$ ion includes the same five bound states (as in the 
$d t \mu e_2$ ion), but in this ion the $[(0,3)_{\mu}; (1^{1}S)_e]$-state is also bound, while the $(1,1)_{\mu}$-state is not weakly-bound. 

In the six-body $a b c \mu e_2$ quasi-molecules situation is similar, but these six-body quasi-molecules are the neutral two-center systems and the total number of bound states in 
such systems can be very large (usually dozens and many dozens of bound states). In the first approximation we can classify these bound states by indicating the quantum $(\nu, 
\ell)_{\mu}$ numbers for the three-body muonic quasi-nucleus, i.e. for the $p d \mu, p t \mu, d t \mu$ and other muonic ions. We also need to add to this $(\nu, 
\ell)_{\mu}$-notation the set of quatum numbers which desribe the two bound electrons. These two electrons form the electron shell in this quasi-molecule. Finally, one obtains the 
notations such as, e.g., $[(\nu, \ell)_{\mu}; ({}^{1}\Sigma^{+}_e)], [(\nu, \ell)_{\mu}; ({}^{1}\Pi)_e]$, etc mentioned above. However, to designate the actual bound state in the 
whole six-body $a b c \mu e_2$ quasi-molecule we also need to indicate the `vibrational' ($v$) and `rotational' ($J(K)$) quantum numbers \cite{LLQ}. These quantum numbers describe 
the quasi-classical motion of two heavy nuclei in the linear two-center $a b c \mu e_2$ quasi-molecule. Below, the pair of quasi-classical quantum numbers $v$ and $J(K)$ are used 
as the doule index of the bound state. Thus, a bound state in the six-body $a b c \mu e_2$ quasi-molecule is designated by the following notation $[(\nu, \ell)_{\mu}; 
({}^{1}\Sigma^{+})_e ]_{v J}$, or $[(\nu, \ell)_{\mu}; ({}^{1}\Pi)_e ]_{v J}$, etc. 

The classification system of bound states in muon-containing five- and six-body systems developed here allows one to designate any bound state in the five- and six-body Coulomb 
systems with unit charges and can succesfully be applied to all five-body $a b \mu e_2$ ions and six-body $a b c \mu e_2$ quasi-molecules. Now, we can describe the bound states in 
these systems. However, our notation $a b c \mu e_2$ used in this study for the six-body $a b c \mu e_2$ systems is, in fact, too restrictive, since it always means that the condition 
$M_a \le M_b \le M_c$ should be obeyed for the masses of the three heavy particle. Such a condition excludes a number of actual bound states. If we remove this restriction, then in 
addition to the $a b c \mu e_2$ system we also need to consider the six-body $c a b \mu e_2$ and $b c a \mu e_2$ quasi-molecules. The central muonic quasi-nucleus in these two 
quasi-molecules are the muonic $a b \mu$ and $c a \mu$ (or $a c \mu$) ions, respectively. It is clear that the three-body muonic $b c \mu$ ion has the lowest total energy 
among the $a b \mu, a c \mu$ and $b c \mu$ ions. In other words, the two light $a b \mu$ and $a c \mu$ ions are not trully stable ions inside of the $a b c \mu e_2$ quasi-molecule. 
However, in actual six-body $a b c \mu e_2$ systems the corresponsing transition times of the central $a b \mu$ and $a c \mu$ ions into `stable' $b c \mu$ ion are relatively large and 
comparable with the muon life-tmie $\tau_{\mu} \approx 2.05 \cdot 10^{-6}$ sec. 

Briefly, this means that we need to consider all these six-body $a b c \mu e_2, b a c \mu e_2$ and $c a b \mu e_2$ systems as stable. Finally, the total number of different series 
of bound electron states increases significantly. The additional series of bound states are called the exchange series. Such series of bound states can be observed in the both 
non-symmetrical $p d t \mu e_2$ and partially symmetrical $a b b \mu e_2$ and $a a b \mu e_2$ six-body quasi-molecules (there are seven similar six-body quasi-molecules). However, 
such exchange series of bound states do not exist in the five-body $a b \mu e_2$ ions and in the truly symmetric $p p p \mu e_2, d d d \mu e_2$ and $t t t \mu e_2$ six-body systems. 
In reality, the exchange series are combined with the regular (or fundamental) series of bound states. The total number of bound states in the six-body $a b c \mu e_2$ quasi-molecules 
can be very large. For instance, in the $p d t \mu e_2$ quasi-molecule the total number of different series of bound electronic states (fundamental + exchange) equals nine, while in 
the partially symmetric $d d t \mu e_2$ quasi-molecule such a number equals ten. As an example, consider the symmetric  $d d t \mu e_2$ quasi-molecule. In this sytem the three-body 
muonic ion can be either $d d \mu$, or $d t \mu$. Each of these muonic ions has five bound states. This leads to the ten bound state series in the $d d t \mu e_2$ system. Analogously,
one of the nuclei in the $p d t \mu e_2$ quasi-molecule can be either $p d \mu$, or $p d \mu$, or $d t \mu$ (this ion is truly stable). Each of the two three-body $p d \mu$ and 
$p d \mu$ ions has two bound states, while the $d t \mu$ ion has five bound states.

To conclude this Section we note that the weakly-bound (1,1)-states in the three-body muonic $d d \mu$ and $d t \mu$ ions (or three-body `quasi-nuclei' in this study) have a 
pre-dissociation structure. Briefly, this means that in the (1,1)-state of the $d t \mu$ muonic ion the nucleus of deuterium $d$ moves at very large distance from the neutral 
two-body quasi-nucleus $t \mu$. The actual $d - (t \mu)$ distance is very large and quite comparable with the $p - (t \mu)$ distance and electron-nucleus distances. The two electrons 
in the six-body $p d t \mu e_2$ system move in the Coulomb field of the two heavy nuclei ($p$ and $d t \mu$), but the spatial position of one of these two Coulomb `nuclei' (or 
centers) rapidly oscillates. In other words, the two bound electrons interact with the bare nucleus of protium and quasi-nucleus ($d t \mu)^{+}$, but for relatively short times the 
nucleus of deuterium (fromthis three-body muonic ion ($d t \mu)^{+}$) acts as an additional source of the Coulomb field, while the neutral particle $t \mu$ doen not interact with 
electrons et al. The frequency and amplitude of such oscillations can approximately be evaluated from our knowledge of the internal structure of the $d t \mu$ ion in its (1, 1) 
weakly-bound state. Formally, this means that our classification scheme of bound state spectra developed for the `regular' six-body $a b c \mu e_2$ systems with unit charges becomes 
quite approximate in those cases when one of the three-body quasi-nucleus in such a system (e.g., $d t \mu$) is in the weakly-bound (1,1)-state.           

\section{Conclusion}

We have investigated the bound state spectra in a number of five- and six-body Coulomb systems with unit charges. In this study we considered the five- and six-body systems each 
of which contains two bound electrons, one negatively charged muon $\mu^{-}$ and two (or three) nuclei of hydrogen isotopes, i.e., protium $p$, deuterium $d$ and tritium $t$. It 
is shown that the ground states in such five- and six-body systems are always stable. Stability of these Coulomb systems with unit charges can be explained by their internal 
clusterization, i.e. the formation of the three-body muonic ion which has a very short spatial radius. In the five-body $a b \mu e_2$ ions and six-body $a b c \mu e_2$ 
quasi-molecules such compact three-body muonic ($a b \mu, a c \mu$ and $b c \mu$) ions play the role of quasi-nucleus with the unit electric charge. Analysis of electronic bound 
state properties leads us to the conclusion about similarity between the internal structures of the five-body, two-electron ions $a b \mu e_2$ [or $(a b \mu e_2)^{-}]$ ion and 
two-electron hydrogen ion H$^{-}$. This fact follows from the accurate numerical coincidence observed for a number of basic geometrical and physical properties of the five-body, 
two-electron ions $a b \mu e_2$ and analogous properties of the three-body hydrogen ion H$^{-}$. Such a similarity can be used to predict that each of the five-body $(a b \mu 
e_2)^{-}$ ions has only one bound electron state for each bound muonic state in the $b c \mu e_2$ ion.

We also investigate stability and determine a number of bound state properties of the six-body $a b c \mu e_2$ quasi-molecules. It is shown that the electron structure of each of 
these six-body quasi-molecules is similar to the structure of the two-center, hydrogen H$_2$ molecule. In contrast with the five-body $(a b \mu e_2)^{-}$ ions, each of the six-body 
quasi-molecules $a b c \mu e_2$ many bound states (electron and molecular states) which belong to a number of different series. Classification of these bound states in the 
$a b c \mu e_2$ systems is a complex problem, which can be simplified by considering separation of the bound state spectra of such systems into a few different series, e.g., into 
the fundamental and exchange series of bound states. 

In conclusion, we wish to note again that direct and accurate variational calculations of the bound state spectra in the five- and six-body two-electron muonic systems $a b \mu e_2$ 
[or $(a b \mu e_2)^{-}$] and $a b c \mu e_2$ is a new and significant step to our understanding of the internal structure of these Coulomb few-body systems with unit charges. 
In the future we need  to increase of the overall accuracy of bound state computations of five- and six-body two-electron muonic systems $a b \mu e_2$ and $a b c \mu e_2$. Another aim 
is to determine more bound state properties, including the expectation values of all inter-particle delta-functions, triple delta-functions, etc. These values are needed to predict the 
actual rates of different physical processes and reactions, e.g., the nuclear fusion rate(s) and transition rates, in different five- and six-body two-electron muonic systems. A 
separate future goal is to analyze the five-body $a b \mu e_2$ ions and six-body $a b c \mu e_2$ quasi-molecules which contain either the $dt\mu$ quasi-nucleus, or $dd\mu$ quasi-nucleus 
in their weakly-bound (1,1)-states. These problems are of great theoretical interest, since in these quasi-nucleus one can observe a strong interaction between the electron (or atomic) 
bound states and weakly-bound (1,1)-states in the three-body quasi-nuclei $d d \mu$ and $d t \mu$. In other words, in such few-body systems we cannot separate the electron's motion 
from internal motion in the three-body $d t \mu$ and $d d \mu$ quasi-nuclei even in the first approximation. This means that these two fundamentally different motions must be considered 
together. Classification of the bound states in the six-body $a b c \mu e_2$ systems with one weakly-bound nucleus is another interesting problem.

Finally, it should be mentioned that current numerical accuracy of our computations is sufficient to make important theoretical predictions about the structure and basic properties 
of all five- and six-body systems with unit charges considered in this study. These few-body systems and their bound state properties have never been considered in earlier studies.
Unfortunately, in 1980's reseachers could not use effective optimization methods to constrct highly effective variational wave-functions for the five- and six-body Coulomb systems. 
This explains the well known fact that probabilities of many important processes in the resonance muon-catalized fusion (see below) were determined very approximately. Here we can say  
that our study opens a new avenue in theoretical and experimental investigations of bound states in the few- and many-body Coulomb systems which contain at least one bound three-body 
cluster of a very small spatial radius (systems with the internal clusterization). 

\section{Appendix. Applications.}

Let us consider applications of the five-body $a b \mu e_2$ (or $(a b \mu e_2)^{+}$) muonic ions and six-body $a b c \mu e_2$ quasi-molecules to actual physical problems. First of all, 
it easy to find that all five- and six-body muonic systems investigated in this study are of great interest in a number of applications. Such applications include muon-catalyzed 
nuclear fusion, theoretical development of the complete theory of internal conversion of electromagnetic radiation in cluster few-body systems, etc. Another interesting direction is 
the use of five-body $a b \mu e_2$ and six-body $a b c \mu e_2$ Coulomb systems for approximation of actual atomic and molecular systems. Advantages of this approximation are obvious, 
since all `quasi-nuclear' properties of the central $a b \mu$ (or $b c \mu$) Coulomb three-body cluster can be evaluated to very high numerical accuracy (in contrast with the actual 
atomic nuclei). Let us discuss some of these applications in detail. 

First, consider the muon catalized nuclear fusion which was experimentally discovered in the liquid protium-deuterim mixture at the end of 1950's \cite{Alv}. However, theoretical 
studies of this process begun ten years earlier (a large number of references to earlier papers can be found in \cite{Alv} - \cite{MS}). Note that in \cite{Alv} only the 
`non-resonance' muon catalized nuclear fusion was considered. During such a `non-resonance' nuclear fusion one negatively charged muon $\mu^{-}$ of high energy rapidly slows down 
inside of the liquid (or dense) protium-deuterium mixture (or HD-mixture, for short) and form the quasi-stable muonic atom $p \mu$ in its ground $1s-$state. This muonic quasi-atom 
is a very compact and neutral two-body systems, which freely moves in the cold and dense HD-mixture and directly interacts with the nuclei of surronding molecules. Such an interaction 
with one of the nucleus in the two-nuclei atomic hydrogen/deuterum molecule leads to the formation of few-body molecular ions. This can be desribed by the following reaction
\begin{eqnarray}
    p \mu + {\rm D}_2 = \{[d (p d \mu)] e\}^{+} + e^{-} \; \; \; , \; \label{nonres} 
\end{eqnarray}
where $e^{-}$ is a free electron emitted during this process. Such a `free' electron removes an excess of the energy released in the process Eq.(\ref{nonres}). The following reaction 
of nuclear $(p,d;{}^{3}He,\gamma)-$fusion in the three-body $(p d \mu)^{+}$ ion proceeds relatively slow, since this reaction proceeds with the emission of $\gamma-$quantum ($p d \mu = 
{}^{3}$He + $\gamma$ + 5.494 $MeV$). As is well known from nuclear physics (see, e.g., \cite{Blat}) the nuclar reactions, which involve an emission/absorbtion of $\gamma-$quantum, are 
usually very slow in a few-nucleon systems. This is the main reason why the $(p,d)-$reaction was never used in the muon-catalized nuclear fusion. The nuclear $(d,d)-$ and 
$(d,t)-$reactions in the analogous $\{[p (d d \mu)] e\}^{+}$ and $\{[d (d t \mu)] e\}^{+}$ five-body ions are significantly faster and this explains a great interest to these nuclear 
reactions and deuterium-tritium systems in applications related to the muon-catalized fusion. However, in a number of earlier experiments it was also shown that the crucial time for 
muon-catalized nuclear fusion is determined by the formation of the five-body quasi-molecular ions $\{[d (p d \mu)] e\}^{+}, \{[d (d d \mu)] e\}^{+}, \{[d (d t \mu)] e\}^{+}$, etc. 
These five-body one-electron structures are similar to the hydrogen molecular ion H$_2^{+}$ which has two heavy Coulomb centers. Finally, the total number of observed reactions of 
nuclear fusion (per one muon) was found to be small ($\le 5$) even for deuterium-tritium species. I became clear that to accelerate the muon catalyzed fusion we need to replace the 
slow reaction, Eq.(\ref{nonres}), by some fast, alternative process. In particular, it is neccessary to avoid any electron ionization of the arising $a b c \mu e_2$ system during such 
a process. 

The discovery of an alternative (or resonance) muon-catalized nuclear fusion was formally made in 1959, when the authors \cite{Bel} tried to determine all bound states in the six 
three-body muonic ions $a b \mu$ (where $(a, b) = (p, d, t)$ by using the pure adiabatic two-center approximation. They have shown that 20 states in these muonic molecular ions are well 
bound and determined their total energies \cite{Bel}. However, the authors \cite{Bel} could not prove the boudness of the (1,1)-states in the $d d \mu$ and $d t \mu$ ions, which are very 
weakly-bound. Nevertheless, the authors \cite{Bel} (see also \cite{Zeld}) made first speculations that such weakly-bound (1,1)-states in the $d d \mu$ and $d t \mu$ ion could lead to a 
significant acceleration of the muon-catalyzed nuclear fusion, since with the help of these states the neutral six-body $[a (d d \mu)] e_2$ and $[a (d t \mu)] e_2$ quasi-molecules (where 
$a$ is an arbitrary hydrogen nucleus, e.g. either $p$, or $d$, or $t$) can be formed during one of the following reactions 
\begin{eqnarray}
    t \mu &+& {\rm D}_2 = [d (d t \mu)] e_2 \; \; \; , \; \; \;  d \mu + {\rm D}_2 = [d (d d \mu)] e_2 \; \; \label{res1} \\
    d \mu &+& {\rm T}_2 = [t (d t \mu)] e_2 \; \; \; , \; \; \;  d \mu + {\rm HD} = [p (d d \mu)] e_2 \; \; \label{res2}
\end{eqnarray}
etc. Note that in any of these reactions no free electron is emitted (in contrast with Eq.(\ref{nonres})), i.e. all these reactions proceed with no ionization of the final six-body 
$a b c \mu e_2$ system. Physically this means that the energy which released during formation of the three-body $d d \mu$ and $d t \mu$ ions in their (1,1)-states transfers directly to 
the excitations of vibrational and rotational levels in the final two-center quasi-molecule $[d (d t \mu)] e_2, [d (d d \mu)] e_2$, etc. Briefly, we have an actual resonance between the 
weakly-bound (1,1)-bound states in the three-body $d d \mu$ and $d t \mu$ ions and the corresponding molecular levels in the six-body two-electron quasi-molecules. In other words, in each 
of the reactions from Eqs.(\ref{res1}) - (\ref{res2}) the final six-body $[a (b c \mu)] e_2$ systems can be found in a certain $\mid v J \rangle$ `molecular state', i.e in a bound state 
which has the corresponding vibrational and rotational molecular quantum numbers. It is clear that the energies of such `internal' molecular excitations must be less than $\approx$ 4.5 
$eV$, which is the binding energy of the hydrogen H$_2$ molecule. Otherwise, the resulting two-center quasi-molecule $[a (b c \mu)] e_2$ (where $a$ = $p, d, t$) will break up, or 
dissociate. 

For the first time the resonance muon-catalyzed nuclear $(d,d)-$ and $(d,t)-$fusion was discovered in experiments performed in Dubna (Russia) in the middle of 1960's (see, e.g., \cite{Fus1}, 
\cite{Fus2}). Discussion of these experiments and other references to original paperes can be found, e.g., in \cite{Petrov} and \cite{MS}. The crucial parameter obtained in these earlier 
experiments was the number of catalyzed fusion reaction per one muon which started the chain of reactions in the liquid deuterium, or unimolar DT-mixture.  The current numbers of nuclear 
fusions per one muon is evaluated as 20 - 40 for liquid deuterium (i.e. for the $d d \mu$ ions) and $\approx$ 150 - 180 for unimolar (or 1:1) DT-miture. Measurments for the $(d,t)-$rection
can be found, e.g., in\cite{Balin}, \cite{Breun} (see also references therein and in \cite{Petrov}, \cite{MS}). In other words, the total energy released in the $(d, t; {}^{4}$He$, n$, 
17.590 $MeV$)-reaction can be evaluated as 180 $\times$ 17.590 $\approx 3166.2 MeV \approx 5.07257 \cdot 10^{-3}$ $erg$. This amount of energy seems to be very large, but it is not sufficient 
(not even close) to compensate all possible energy losses (see e.g., \cite{Fro2011} and \cite{MS}). Another serious complication follows from the fact that $\approx$ 80 \% of all released 
energy is transfered by the fast (or 14.1 $MeV$) neutrons which are formed in large numbers during these fusion reactions. The kinetic energies of these fast neutrons cannot easily be utilized 
in the process which proceeds at the density of liquid hydrogen ($\rho \approx$ 0.213 $g \cdot cm^{-3}$). Furthermore, large numbers of fast neutrons released during the nuclear $(d,t)-$fusion 
present a great danger for woring personel. This explains why the muon catalyzed nuclear fusion will never be used for the energy production purposes. Novadays, the resonance muon-catalyzed 
nuclear fusion is considered as an interesting physical phenomenon, which, however, has no meaning for the energy production. Almost all investigations of the 1resonance' muon-catalized nuclear 
were halted approximately 10 - 15 years ago and now this scientific direction is in a `sleeping mode'.     

Another ineresting application of the results of our study is accurate evaluations of the probability of internal conversion of nuclear radiation into the energy of emitted electron, which 
was originally, bound to the same atom(s) and/or molecule(s). In general, the internal conversion probability amplitude is represented as the sum of products of the two following factors 
(see, e.g., \cite{AB}): (a) the nuclear factor, and (b) the electronic factor. Currently, the second (electronic) factor can be computed to high and very high accuracy for an arbitrary, in 
principle, atom/molecule. Such a conclusion is based on a significant progress made recently in accurate atomic and molecular computations. However, the nuclear factors are still determined 
with large uncertainties and fundamental inaccuracies. Finally, in many cases the overall errors in the internal conversion probability amplitudes exceed 10 - 20 \%. For the five-body $a b 
\mu e_2$ (or $(a b \mu e_2)^{+}$) muonic ions and six-body $a b c \mu e_2$ quasi-molecules considered in this study the corresponding `nuclear' (or quasi-nuclear) factor has the same (usually, 
high or very high) numerical accuracy as the electronic factor. This allows one to gain a large number of theoretical and numerical advantages by investigating the process of internal 
conversion of radiation, emitted by the central three-body cluster $(a b \mu)^{+}$, in the five- and six-body muonic systems $a b \mu e_2$ (or $(a b \mu e_2)^{+}$) and $a b c \mu e_2$.

\newpage
\begin{table}[tbp]
   \caption{Total energies in atomic units ($a.u.$) of the ground states in the five-body hydrogen-muonic ions $a b \mu e_2$ (or $(a b \mu e_2)^{-}$) and 
             six-body quasi-molecules $a b c \mu e_2$.}
     \begin{center}
     \begin{tabular}{| c | c | c | c |}
      \hline\hline
 $N$ &  $p d \mu e_2$ & $p t \mu e_2$ & $d t \mu e_2$ \\ 
     \hline
 200 & -106.53525247 & -108.017225101 & -111.88748567 \\
         \hline\hline
 $N$  & $p p \mu e_2$ & $d d \mu e_2$ & $t t \mu e_2$ \\ 
     \hline
 200 & -102.75006453 & -110.34353875 & -113.49917375 \\
       \hline\hline
 $N$  & $p d t \mu e_2$ & $p p d \mu e_2$ & $p p t \mu e_2$ \\ 
     \hline
 200 & -112.47568634 & -107.12465837 & -108.60740196 \\
         \hline 
 $N$ & $d d t \mu e_2$ & $p d d \mu e_2$ & $p t t \mu e_2$ \\ 
     \hline
 200 & -112.47230678 & -110.95256865 & -114.10638563$^{(a)}$ \\
         \hline 
 $N$  & $p p p \mu e_2$ & $d d d \mu e_2$ & $t t t \mu e_2$ \\ 
     \hline
 200 & -103.33607025 & -110.92843415 & -114.08028863 \\
         \hline \hline  
  \end{tabular}
  \end{center}
 ${}^{(a)}$The total energy of the ground bound state of the $d t t \mu e_2$ system is -114.10539729 $a.u.$
  \end{table}
\begin{table}[tbp]
   \caption{The total energies $E$ in atomic units ($a.u.$) of the ground states of the three-body muonic molecular ions $a b \mu$ (or $(a b \mu)^{+}$).
            Numerical values have been taken from \cite{Fro2011}. All particle masses are exactly the same as they used in this study.} 
     \begin{center}
     \begin{tabular}{| c | c | c | c |}
      \hline\hline
 ion &  $p p \mu$ & $d d \mu$ & $t t \mu$ \\ 
     \hline
 $E$ & -102.2235035785787 & -109.8169263959981 & -112.9728490317648 \\
         \hline\hline
 ion &  $p d \mu$ & $p t \mu$ & $d t \mu$ \\ 
     \hline
 $E$ & -106.0125270695158 & -107.4947026128185 & -111.3643469153818 \\ 
         \hline \hline  
  \end{tabular}
  \end{center}
  \end{table}
%
%
 \begin{table}[tbp]
   \caption{The expectation values of a number of properties (in $a.u.$) of the ground states in some five-body $a b \mu e_2$ (or $(a b \mu e_2)^{-}$) 
            ions and six-body $a b c \mu e_2$ quasi-molecules.}
     \begin{center}
     \begin{tabular}{| c | c | c | c | c |}
       \hline\hline          
 ion/molecule  & $\langle r^{-2}_{d \mu} \rangle$ & $\langle r^{-1}_{d \mu} \rangle$ & $\langle r_{d \mu} \rangle$ & $\langle r^2_{d \mu} \rangle$ \\
     \hline
   $p d \mu e_2$ & 38966.70 & 132.5671 & 0.01185483 & 0.00018781 \\
  
   $d d \mu e_2$ & 48962.04 & 150.6274 & 0.01025259 & 0.00013908 \\

 $p d t \mu e_2$ & 47950.20 & 149.4190 & 0.01024187 & 0.00013845 \\ 
     \hline\hline     
 ion/molecule  & $\langle r^{-2}_{d e} \rangle$ & $\langle r^{-1}_{d e} \rangle$ & $\langle r_{d e} \rangle$ & $\langle r^2_{d e} \rangle$ \\
     \hline
   $p d \mu e_2$ & 1.100883 & 0.6863477 & 2.569325 & 9.871617 \\
  
   $d d \mu e_2$ & 1.110315 & 0.6841402 & 2.326188 & 10.98206 \\ 

 $p d t \mu e_2$ & 1.489877 & 0.8835500 & 1.606683 & 3.251662 \\
     \hline\hline     
 ion/molecule  & $\langle r^{-2}_{ee} \rangle$ & $\langle r^{-1}_{ee} \rangle$ & $\langle r_{ee} \rangle$ & $\langle r^2_{ee} \rangle$ \\
     \hline
   $p d \mu e_2$ & 0.161285 & 0.320110 & 4.12698 & 20.88617 \\
  
   $d d \mu e_2$ & 0.157155 & 0.313956 & 4.29985 & 23.26188 \\

 $p d t \mu e_2$ & 0.506331 & 0.573921 & 2.22105 & 5.857835 \\
     \hline\hline     
 ion/molecule  & $\langle \frac12 {\bf p}^2_{e} \rangle$ & $\langle \frac12 {\bf p}^2_{\mu} \rangle$ & $\langle \frac12 {\bf p}^2_{p} \rangle$ & $\langle \frac12 {\bf p}^2_{d} \rangle$ \\ 
     \hline
   $p d \mu e_2$ & 0.2633550 & 19683.348 & 15710.384 & 11997.316 \\
  
   $d d \mu e_2$ & 0.2635159 & 20880.859 & --------  & 16204.142 \\ 

 $p d t \mu e_2$ & 0.9912942 & 21485.471 & 23781.425 & 20912.580 \\
      \hline\hline
  \end{tabular}
  \end{center}
 ${}^{(a)}$The $\langle \frac12 {\bf p}^2_{t} \rangle$ expectation value for the $p d t \mu e_2$ quasi-molecule is $\approx$ 21149.609.
   \end{table}
%

 \begin{table}[tbp]
   \caption{The expectation values of a number of properties (in $a.u.$) of the ground states in the ${}^{\infty}$H$^{-}$ ion and ${}^{1}$H$_{2}$ molecule.
            The notation $e$ stands for the electron(s), while $N$ designates the heavy nucleus. The total energy of the ${}^{\infty}$H$^{-}$ ion is $\approx$ 
            -0.52775101 65443771 9659085(10) $a.u.$, while the total energy of the ${}^{1}$H$_{2}$ molecule is $\approx$ -1.1638958(25) $a.u.$}
     \begin{center}
     \scalebox{0.85}{%
     \begin{tabular}{| c | c | c | c | c |}
       \hline\hline          
 ion/molecule  & $\langle r^{-2}_{N e} \rangle$ & $\langle r^{-1}_{N e} \rangle$ & $\langle r_{N e} \rangle$ & $\langle r^2_{N e} \rangle$ \\
     \hline
           H$^{-}$ & 1.11666282452542572 & 0.6832617676515272224 & 2.7101782784444203653 & 11.913699678051262274 \\
     \hline
   ${}^{1}$H$_{2}$ & 1.572275 & 0.9014961 & 1.575480 & 3.147748 \\
     \hline    
  ion/molecule  & $\langle r^{-2}_{e e} \rangle$ & $\langle r^{-1}_{e e} \rangle$ & $\langle r_{e e} \rangle$ & $\langle r^2_{e e} \rangle$ \\
     \hline
           H$^{-}$ & 0.15510415256242466 & 0.311021502214300052 & 4.4126944979917277211 & 25.202025291240331897 \\
    \hline
   ${}^{1}$H$_{2}$ & 0.5046369 & 0.5791711 & 2.2018442 & 5.8074634 \\
         \hline 
  molecule  & $\langle r^{-2}_{N N} \rangle$ & $\langle r^{-1}_{N N} \rangle$ & $\langle r_{N N} \rangle$ & $\langle r^2_{N N} \rangle$ \\
          \hline 
   ${}^{1}$H$_{2}$ & 0.496476 & 0.6993974 & 1.450728 & 2.134826 \\  
      \hline\hline
  \end{tabular}}
  \end{center}
 ${}^{(a)}$The $\langle \frac12 {\bf p}^2_{e} \rangle$ expectation value for the ${}^{\infty}$H$^{-}$ ion (computed with the same wave function) 
           is $\approx$ 0.2638755082721885983. 
   \end{table} 
\end{document}